# Three-Photon Absorption Spectra and Bandgap Scaling in Direct-Gap Semiconductors

**SEPEHR BENIS[1], CLAUDIU M. CIRLOGANU[2], NICHOLAS COX[1], TRENTON ENSLEY[3], HONGHUA HU[4], GERO NOOTZ[5], PETER D. OLSZAK[6], LAZARO A. PADILHA[7], DAVORIN PECELI[8], MATTHEW REICHERT[9], SCOTT WEBSTER[1], MILTON WOODALL[10], DAVID J. HAGAN[1, \*], AND ERIC W. VAN STRYLAND[1]**

[1]*CREOL, The College of Optics and Photonics, University of Central Florida, Orlando, FL 32816, USA*
[2]*Currently with Lumileds, San Jose, CA 95131, USA*
[3]*Currently with CCDC - U.S. Army Research Laboratory, Sensors and Electron Devices Directorate, Adelphi, MD 20783, USA*
[4]*Currently with MKS Instruments, Portland, OR 97229, USA*
[5]*Currently with Division of Marine Sciences, School of Ocean Science and Engineering, University of Southern Mississippi, Stennis Space Center, MS 39529, USA*
[6]*Currently with Mount Wachusett Community College, Gardner, MA 01440, USA*
[7]*Currently with Instituto de Fisica "Gleb Wataghin", Universidade de Campinas, Campinas, 13083-970 Sao Paulo, Brazil*
[8]*Currently with Extreme Light Infrastructure – Beamlines, FZU AS CR, v.v.i., 252 41 Dolni Brezany, Czech Republic*
[9]*Currently with CACI International Inc., Florham Park, NJ 07932, USA*
[10]*Currently with Leonardo DRS, Dallas, TX 75251, USA*
*\*Hagan@creol.ucf.edu*

**Abstract:**
This paper presents three-photon absorption (3PA) measurement results for nine direct-gap semiconductors, including full 3PA spectra for ZnSe, ZnS, and GaAs. These results, along with our theory of 3PA using an 8-band Kane model (4 bands with double spin degeneracy), help to explain the significant disagreements between experiments and theory in the literature to date. 3PA in the 8-band model exhibits quantum interference between the various possible pathways that is not observed in previous 2-band theories. We present measurements of degenerate 3PA coefficients in InSb, GaAs, CdTe, CdSe, ZnTe, CdS, ZnSe, ZnO, and ZnS. We examine bandgap, $E_g$, scaling using 2-band tunneling and perturbation theories that show agreement with the predicted $E_g^{-7}$ dependence; however, For those semiconductors for which we measured full 3PA spectra, we observe significant discrepancies with both 2-band theories. On the other hand, our 8-band model shows excellent agreement with the spectral data. We then use our 8-band theory to predict the 3PA spectra for 15 different semiconductors in their zinc-blende form. These results allow prediction and interpretation of the 3PA coefficients for various narrow to wide bandgap semiconductors.

## 1. INTRODUCTION

In the past decades, three-photon absorption (3PA) has shown both fundamental research importance and technological implications in many nonlinear optical (NLO) applications [1-8]. In organic optoelectronics, 3PA can be used as an excitation to observe efficient stimulated emission and frequency upconversion fluorescence emission [2-4]. 3PA can contribute to high-harmonic generation in semiconductor materials [9, 10] or may compete with two-photon lasing in direct gap semiconductors [11, 12]. Additionally, 3PA in semiconductor quantum dots and nanocrystals has attracted major interest for applications in bio-labeling and imaging agents due to the possibility of using longer excitation wavelengths to achieve deeper penetration depths for super-resolution imaging [1, 13-15]. Similarly, 3PA allows the possibility of using light sources in the telecommunication range for optoelectronic applications in wide-bandgap materials such as ZnO [5, 13, 16, 17]. Also, 3PA in semiconductors can be a limiting factor in all-optical switching applications below half the bandgap [8, 18]. Thus, accurate modeling and experimental verification of 3PA in semiconductors is of great importance for design and characterization of various NLO devices.

While theory and experiment for two-photon absorption (2PA) in direct-gap semiconductors are in excellent agreement [19-25], previous comparisons for 3PA have yielded mixed results [16, 20, 26-29]. Two convenient scaling rule theories, assuming only two parabolic bands, were developed by Brandi and de Araujo using Keldysh's tunneling approach in 1983 [26, 29], and by Wherrett using perturbation theory in 1984 [20]. These theories, after reformulating the effective mass, give identical results for 2PA. They also yield similar results for 3PA, differing only in the spectral response and an overall multiplicative constant. Both theories predict that the 3PA coefficient, $\alpha_3$, scales with the bandgap energy, $E_g$, as $E_g^{-7}$ [20, 26, 29]. They find a spectral dependence of $(3\hbar\omega/E_g - 1)^X / (3\hbar\omega/E_g)^9$, where tunneling theory yields $X = 5/2$ and perturbation theory gives $X = 1/2$. Arbitrary constant scaling factors have been applied to fit measurement results to theory, as was done for 2PA [19, 21]. For example, Brandi and de Araujo scaled the calculated 2PA coefficient by an arbitrary factor of six to match data for CdS [29].

Here, we employ a Kane 8-band model (4 bands with double spin degeneracy) [30] to predict the 3PA spectra of 15 direct-gap semiconductors and present Z-scan measurements of the full 3PA spectra of ZnS, and GaAs. In addition, we compare our previous measurement of the 3PA spectrum of ZnSe [27, 28] to the 8-band calculations. Quantum interference between the various pathways available for valence to conduction transitions leads to a more richly featured 3PA spectrum compared to that obtained from simple 2-band models. In particular, 3PA spectra show a local peak at photon energies just above $E_g/3$ followed by a relatively flat spectral response through the turn-on of 2PA at $E_g/2$. A second local maximum may appear between $E_g/3$ and $E_g/2$ depending on the spin-orbit splitting. The data and theory for these three materials are in excellent agreement for the shape of the spectra; however, as mentioned, an overall scaling factor of 2 to 4 is needed to match the overall magnitude.

We also performed measurements of 3PA in CdS, ZnO, ZnTe, CdTe, CdSe, and InSb at selected wavelengths. We show these data along with a few selected literature measurements to demonstrate the $E_g^{-7}$ scaling [6, 16, 31-34]. However, in order to plot this bandgap scaling, a constant spectral shape must be assumed. This assumption leads to discrepancies for different spectral measurements of a single material since spectral data closely follow the spectra predicted by the 8-band theory. These spectra change from material to material, e.g. the spin-orbit splitting differs for each material. The 8-band theory calculates the 3PA spectra by separately predicting the contributions of all transitions where the heavy-hole (HH), light-hole (LH), and split-off (SO) bands are the initial states. The agreement between our measured spectra and theory for three semiconductors allows us to accurately establish references for predicting and interpreting 3PA spectra for 12 additional zinc-blende semiconductors. Reasonable estimates for materials with different structures could also be made. The role of spin-orbit coupling is crucial in understanding the quantum interference and helps to understand the spectral shape of the 3PA.

## 2. EXPERIMENTAL PROCEDURES

We measured 3PA coefficients using several different pulsed laser systems by employing one of two methods: direct transmittance measurements [21, 35] (labeled "T" in Table 1) and open-aperture Z-scans [36, 37] (labeled "Z" in Table 1). Our Z-scan data in Table 1 was taken with a tunable femtosecond source (Clark MXR 2010 pumped TOPAS-C optical parametric generator amplifier- OPG/A). Details of the experimental setup are given in [27, 28], and an example of data fitting is shown in Fig. 1. The literature data for $As_2Se_{1.5}S_{1.5}$ glass was taken using the same laser system and analyzed in the same way as the femtosecond data in this work [38]. 3PA coefficients for ZnO, ZnSe and CdS at 1064 nm were found by measuring transmission ($T$) versus irradiance ($I$) using 30 picosecond FWHM pulses from a passively modelocked Quantel Nd:YAG laser. The transmission, T, was measured as a function of input irradiance. These values were used to plot $1/T^2$ versus $I^2$ which theoretically yields straight lines for constant irradiance in time and space with a slope determining the 3PA coefficient. However, integrals over the Gaussian temporal and spatial coordinates give a slight downward curvature to these lines. The theoretical fits to these curves at low irradiance yields $\alpha_3$ [39]. At high irradiance, effects of FCA increase the losses as discussed below. Measurements on InSb at 12 μm were taken using pulses created via difference frequency generation (DFG) of picosecond pulses from a hybrid modelocked Ekspla PL 2143 laser system.

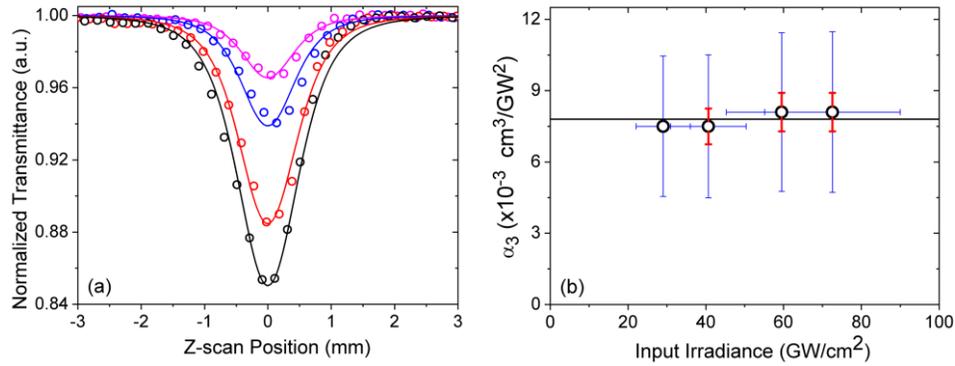

Fig. 1. (a) Z-scans at several peak irradiances in ZnSe at 1050nm (29 GW/cm² pink, 41 GW/cm² blue, 60 GW/cm² red, and 72 GW/cm² black) along with fits (solid lines) using the values for $\alpha_3$ shown in (b). (b) $\alpha_3$ values obtained from the data in (a). The error bars (blue) are absolute while the errors (red) relative to the lowest irradiance Z-scan are considerably smaller. The average $\alpha_3 = (7.8 \pm 3) \times 10^{-3} cm^3/GW^2$ while considering there is no free-carrier absorption.

Measurement accuracy of 3PA coefficients decreases with increasing pulsewidth due to the additional buildup of 3PA-excited carriers. The subsequent free-carrier absorption (FCA) can dominate 3PA for picosecond and nanosecond pulses, and the accompanying free-carrier refraction makes it difficult to collect all the transmitted energy [40, 41]. For pulses less than ~100 ps, recombination can be ignored so that pulse propagation is simply modeled by considering only 3PA and FCA [39]. In this model, the change in irradiance induced by 3PA and FCA is formulated as:

$$\frac{dI}{dz} = -\alpha_3 I^3 - \sigma_{FCA} N I \text{ and } \frac{dN}{dt} = \frac{\alpha_3 I^3}{3\hbar\omega}, \qquad (1)$$

where $\alpha_3$ is the 3PA coefficient, $\sigma_{FCA}$ is the FCA cross-section, and $N$ is the density of carriers excited via 3PA. These equations are solved numerically for both $\alpha_3$ and $\sigma_{FCA}$ to fit the picosecond Z-scan measurements [42]. In all the measurements reported here, we found FCA to be negligible except for the picosecond Z-scan measurements of GaAs and InSb (80 K). Our "T" measurements also used $\sigma_{FCA} = 0$.

Data from other references are included in Table 1, distinguished by italicized material names. Material parameters used in the theory discussed in the next section are also given along with the theoretical $\alpha_3$ calculated by Wherrett's model. The Keldysh model gives similar results but with somewhat larger discrepancies as explained in section 3. Broad spectral data taken for ZnS, ZnSe, and GaAs are shown in section 4. Values for these spectra are not listed in the table but are included in Fig. 2.

Table 1. 3PA coefficients (experimental data, $\alpha_3^{exp}$, and values from Eq. 2, Wherrett) with relevant material parameters. Materials in italics are taken from the references listed. All $\alpha_3$ values are in $cm^3/GW^2$. τp is the pulsewidth FWHM. The experimental method is either direct transmittance (T) or open-aperture Z-scan (Z). The band parameters used are listed in Table 2.

| Material | $E_g$ [eV] | $E_p$ [eV] | $n$ | $\lambda$ [μm] | $\alpha_3^{exp}$ | Wherrett | $\tau_p$ | Method |
|---|---|---|---|---|---|---|---|---|
| *ZnS* [a] *[5]* | 3.54 | 20.4 | 2.3 | 0.8 | 0.0017 | 0.00130 | ~100fs | Z |
| *ZnO* [a] *[5]* | 3.27 | 28.2 | 1.95 | 0.9 | 0.0054 | 0.00506 | ~100fs | Z |
| ZnO | 3.27 | 28.2 | 1.94 | 1.06 | 0.022 | 0.0112 | 30ps | T |
| ZnSe | 2.67 | 24.2 | 2.48 | 1.06 | 0.015 | 0.0097 | 30ps | T |
| CdS | 2.42 | 21 | 2.34 | 1.06 | 0.015 | 0.0090 | 30ps | T |
| CdS | 2.42 | 21 | 2.34 | 1.2 | 0.011 | 0.022 | 150fs | Z |
| ZnTe | 2.28 | 19.0 | 2.8 | 1.2 | 0.02 | 0.0196 | 150fs | Z |
| CdSe | 1.90 | 20.0 | 2.5 | 1.5 | 0.24 | 0.085 | 150fs | Z |
| *CdSe [33]* | 1.90 | 20.0 | 2.5 | 1.54 | 0.13 | 0.103 | ~100fs | T |
| *As$_2$Se$_{1.5}$S$_{1.5}$* [b] *[38]* | 1.74 | 21 | 2.7 | 1.55 | 0.055 | 0.086 | 150fs | Z |
| *Al$_{0.18}$Ga$_{0.82}$As [32]* | 1.65 | 21 | 3.34 | 1.55 | 0.05 | 0.045 | ~1ps | T |
| CdTe | 1.44 | 20.7 | 2.7 | 1.75 | 1.2 | 0.19 | 150fs | Z |
| *GaAs* [a] *[6]* | 1.42 | 28.9 | 3.4 | 2.3 | 0.35 | 0.82 | 100fs | Z |
| *InAs [31]* | 0.354 | 21.5 | 3.42 | 9.54 | 1,000 | 11,000 | ~120ps | Z |
| *InSb[34]* | 0.174 | 23.3 | 3.95 | 10.6 | 200,000 | 18,400 | 45ps | T |
| InSb (at 80K) | 0.228 | 21 | 3.95 | 12 | 25,000 | 46,000 | 10ps | Z |

[a]From the spectra given, the single datum is taken from the region of the graph where the dispersion shown was low.
[b]The bandgap of As$_2$Se$_{1.5}$S$_{1.5}$ (1.74 eV) is determined from the linear transmittance spectrum [38].

## 3. BANDGAP SCALING OF THREE-PHOTON ABSORPTION

While the 8-band model exhibits the same bandgap scaling as the 2-band models, the spectral shape is considerably different making direct comparisons of this scaling problematic. Therefore, we look at using Wherrett's model [20] to demonstrate bandgap scaling. As it turns out, its spectral dependence yields a better fit to the data than that of Brandi and de Araujo. However, as we show below, neither model adequately describes the 3PA spectra. Using a 2-parabolic band model, Wherrett's theory gives the 3PA coefficient as:

$$\alpha_3(\omega) = \frac{K_3 E_p^{3/2}}{n^3 E_g^7} F_3\left(\frac{\hbar\omega}{E_G}\right) \text{ where } F_3(x) = \frac{(3x-1)^{1/2}}{(3x)^9}, \quad (2)$$

where $n$ is the refractive index and $E_p$ is the Kane energy. The Kane energy is a measure of the coupling strength between valence and conduction bands, and is around 21 eV for many semiconductors [20, 21, 30]. Both $E_g$ and $E_p$ have units of eV in Eq. 2, and the constant $K_3 \cong 25.1 \ eV^{11/2} cm^3/GW^2$ is obtained from the 2-band theory of Wherrett as estimated by Woodall [20, 30, 39].

Wherrett's perturbative approach assumes dominance of so-called allowed-allowed-allowed transitions, for which each transition is symmetry-allowed at the Brillouin zone center [20]. On the other hand, the tunneling theory of Brandi and de Araujo gives equivalent results to perturbation theory considering only allowed-forbidden-forbidden transitions [29]. In this case, the forbidden steps are intraband "self" transitions. These two approaches give identical 2PA coefficients because both theories only permit allowed-forbidden combinations in a 2-band model. This agreement no longer holds for 3PA, since the triply-allowed transition pathway is ignored by Brandi and De Araujo and the allowed-forbidden-forbidden paths are ignored by Wherrett.

Although the data in Table 1 cover many different semiconductors and wavelengths, we can rescale the 3PA coefficients onto a single universal plot by taking advantage of the theory's explicit bandgap dependence; the remaining spectral dispersion is identical for each material in Wherret's model and can be divided out. Labelling the experimentally measured 3PA coefficients as $\alpha_3^{exp}$, we can isolate the bandgap dependence of Eq. 2 by scaling as:

$$\alpha_3^{scaled}(E_g) \equiv \alpha_3^{exp} \cdot \frac{n^3}{E_p^{3/2} F(x)}. \tag{3}$$

The result should be close to the theoretical value of $K_3 \times E_g^{-7}$. Thus, if we plot the experimentally scaled $\alpha_3^{scaled}$ versus $E_g$ we expect a simple inverse-seventh power dependence.

The 3PA coefficients from Table 1 are scaled according to Eq. 3 and plotted versus bandgap energy in Fig. 2. We find good overall agreement in the scaling but with notable exceptions. For example, the literature datum for InAs is considerably smaller than predicted. This error may arise because long-wave infrared experiments are much more difficult than those at shorter wavelengths, and the 120 ps pulses used cause further complications from FCA [31]. Also, the value for CdTe is six times larger than theory. Data points at multiple wavelengths are available for GaAs, ZnSe, and ZnS, and their vertical spread demonstrate the inadequacy of Wherrett's spectral function. If Eq. 2 provided an accurate depiction of the frequency scaling, then dividing them out as per Eq. 3 would ensure that all data for a given material lie at the same point. These observations led us to investigate the theory of 3PA using the more sophisticated Kane 8-band model for zinc-blende semiconductors and to measure the full 3PA spectrum of ZnSe, ZnS, and GaAs [27, 28]. The theory, developed using third-order perturbation theory, is detailed in Appendix A.

It is important to note that both 2-band models ignore forbidden interband transitions between hole states that are accounted for in the present work; interactions between HH, LH, and SO bands are a crucial component of the 3PA dispersion. Furthermore, including all possible pathways leads to interference effects in the perturbation expansion, so that 3PA cannot even be well-modeled by summing scaled versions of the Wherrett and tunneling theories.

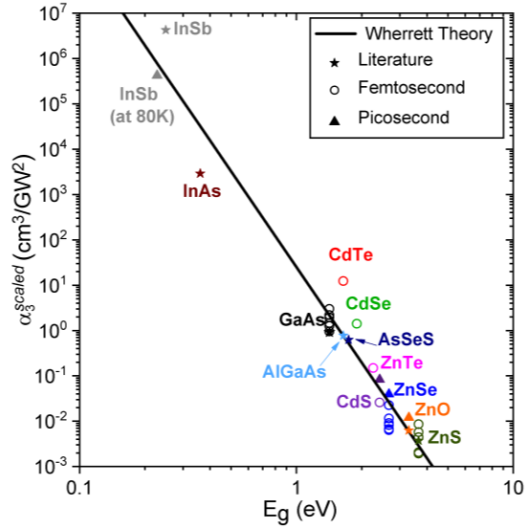

Fig. 2. Scaled 3PA ($\alpha_3^{scaled}$) data from literature in Table 1 (stars), picosecond measurements (Δ's) [39] and Δ for InSb at 80 K, and femtosecond data (circles) compared to Wherrett's theory (solid line). Values taken from Table 1, and femtosecond spectra for ZnSe and ZnS (every other datum graphed), and GaAs (all fs data graphed) see Figs. 3-5.

## 4. SPECTRAL DEPENDENCE OF THREE-PHOTON ABSORPTION

A detailed theoretical approach including all possible transitions for the 8-band model was developed and shown to agree with the measured spectral data for ZnSe [27, 28]. Appendix A gives details on that calculation. A comparison of the theoretically predicted 3PA spectrum for ZnSe (modified from [27, 28]) is shown in Fig. 3(a) as a function of the three-photon energy sum. The theory is scaled by a factor of 1.9 to better fit the data. The results of similar experiments using femtosecond pulses on ZnS are shown in Fig 3(b), where the theory curve is scaled by 2.3. The scaling factor is found via minimizing the sum of square of residuals.

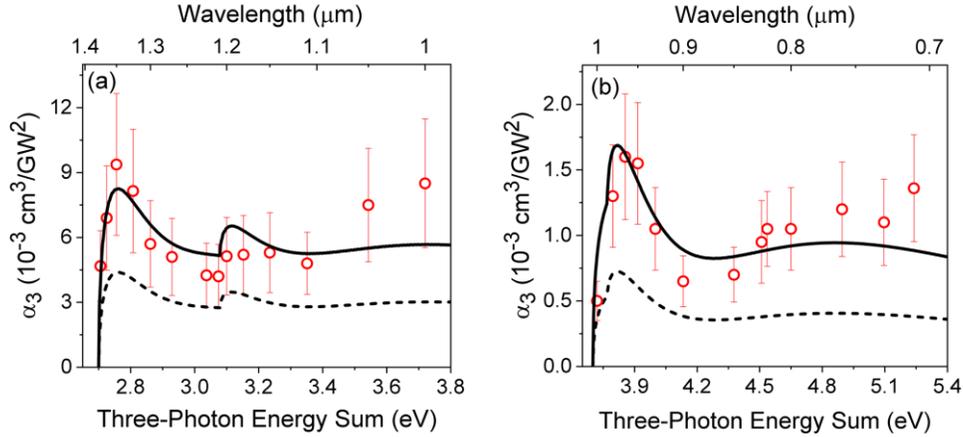

Fig. 3. Measured data (circles) for 3PA in (a) ZnSe (taken from [27, 28]) and (b) ZnS obtained from Z-scans, compared to calculated spectra from the 8-band model (solid black line). Scaling the calculated spectra by factors of 1.9 and 2.3, respectively (red curves), provides the best agreement with experimental data. Band parameters used for theoretical calculation are provided in Table 2.

These data agree with all the salient features of the new theoretical 3PA spectra. The SO band contribution in ZnS is not resolved because it lies within the 3PA peak while it is clearly apparent near 1.2 $\mu m$ in the model for ZnSe. This behavior is expected because the SO energy for ZnS, $\Delta = 0.06$, is small compared to $\Delta = 0.42$ for ZnSe [43].

Having obtained agreement for the wide-gap semiconductors ZnS and ZnSe, we chose to study the 3PA spectrum of the commonly used narrower bandgap semiconductor GaAs. Published data for the 3PA spectrum includes only four wavelengths, making it difficult to convincingly support any theory [6]. We measured the GaAs 3PA spectrum with both picosecond and femtosecond systems. Picosecond measurements of GaAs were made with an Ekspla OPG/A system pumped by a hybrid modelocked Ekspla PL 2143 laser generating pulsewidths of approximately 15 ps (FWHM) in the spectral range of 420 to 2300 nm. FCA contributes significantly to the nonlinear absorption at this pulsewidth, so Z-scans were performed at several different pulse energies to fit the 3PA and FCA coefficients using Eq. 1. The 3PA coefficients obtained are shown in Fig. 4(a) (blue), while the FCA cross sections are shown in Fig. 4(b). The FCA values are in qualitative agreement with those reported in Ref. [44].

For femtosecond measurements, we narrowed the laser pulse spectrum with band-pass filters at each wavelength to better resolve the 3PA edge and peak. Again performing Z-scans, we obtained the data (red) shown in Fig. 4(a). Taking data at several pulse energies did not reveal any FCA, as expected for sub-picosecond pulsewidths. In the end, we find relatively good agreement between picosecond and femtosecond 3PA measurements. The error bars in both graphs indicate relative errors between measurements, and the overall absolute errors are greater by a factor of 2.

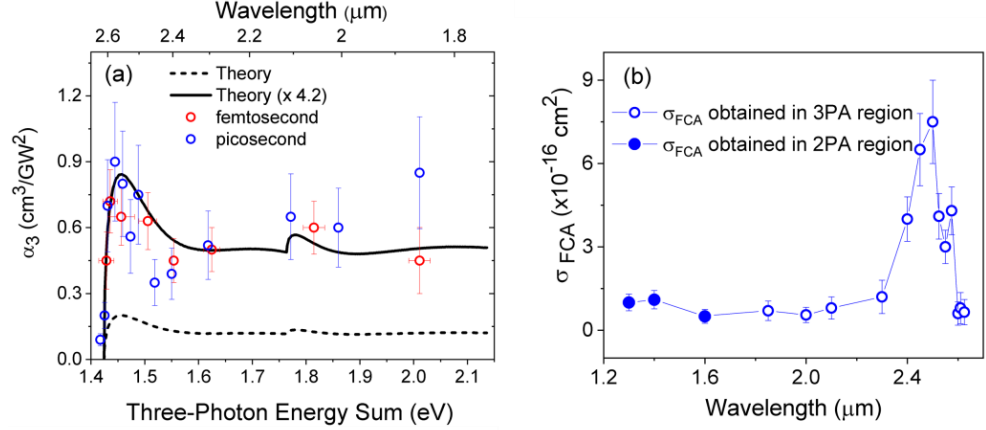

Fig. 4. (a) Measured data for 3PA in GaAs obtained from Z-scans compared to the calculated spectrum from the 8-band model (solid black line). Scaling the calculated spectrum by a factor of 4.2 (red curve), provides the best agreement with experimental data. Band parameters used for the theoretical calculation are provided in Table 2. Blue open circles are experimental data taken with ~150 fs pulses. Purple stars are data for the ~ 10 ps pulses. (b) The FCA cross section from the 2-parameter fit for 3PA and the FCA.

## 5. PREDICTED THREE-PHOTON ABSORPTION SPECTRA AND SPIN-ORBIT COUPLING

The Kane 8-band model, detailed in Appendix A, is used to predict the 3PA spectra of zinc-blende semiconductors. The only parameters needed are the bandgap, Kane energy, and SO energy, which are compiled in Table 2 for various zinc-blende direct-gap semiconductors. We include a separate column for the ratio $\Delta/E_g$ because it turns out to be an important factor in the 3PA spectral shape. This ratio spans from the order of unity in narrow-gap materials to $10^{-2}$ to $10^{-3}$ in wide-gap materials.

Table 2. Band-structure parameters for semiconductors in their zinc-blende form at 300 K (except ZnO, at 4.2 K) used for modeling of 3PA. The first three rows indicate materials where the SO band energy is relatively large ($\Delta/E_g > 1$) and the last five rows indicate materials with a the SO band energy is relatively small ($\Delta/E_g \ll 1$) as discussed in detail below.

|  | Material | $E_g [eV]$ | $\Delta [eV]$ | $E_p [eV]$ | $\Delta/E_g$ | Reference |
|---|---|---|---|---|---|---|
|  | InSb | 0.174 | 0.81 | 23.3 | 4.65 | [30, 45] |
| $\Delta/E_g > 1$ | InAs | 0.354 | 0.39 | 21.5 | 1.10 | [45] |
|  | GaSb | 0.726 | 0.80 | 22.0 | 1.10 | [46, 47] |
|  | InP | 1.344 | 0.11 | 20.4 | 0.08 | [48, 49] |
|  | GaAs | 1.42 | 0.34 | 28.9 | 0.24 | [45, 50] |
|  | CdTe | 1.44 | 0.91 | 20.7 | 0.63 | [43] |
|  | CdSe | 1.90 | 0.42 | 20.0 | 0.22 | [21, 43] |
|  | ZnTe | 2.28 | 0.92 | 19.0 | 0.40 | [21, 43] |
|  | CdS[a] | 2.42 | 0.07 | 21.0 | 0.029 | [51, 52] |
|  | ZnSe | 2.67 | 0.42 | 24.2 | 0.16 | [21, 53] |
|  | GaN | 3.24 | 0.017 | 25.0 | 0.005 | [45] |
|  | ZnO[b] | 3.44 | 0.0087 | 21.0 | 0.002 | [54] |
| $\Delta/E_g \ll 1$ | ZnS | 3.54 | 0.06 | 20.4 | 0.017 | [21, 53] |

| | | | | | |
|---|---|---|---|---|---|
| AlP | 3.56 | 0.07 | 17.7 | 0.02 | [45] |
| AlN | 4.90 | 0.02 | 27.1 | 0.001 | [43, 45] |

[a]Table 2 is sorted based on the values of $E_g$. However, note that CdS belongs to the $\Delta/E_g \ll 1$ section.
[b]8-band model parameters for ZnO are given for T = 4.2 K (Liquid helium temperature).

Fig. 5 gives our predictions of the 3PA spectra for several zinc-blende semiconductors. We explicitly show the contributions from transitions that start in each of the three valence bands: HH, LH, and SO. These bands serve as the initial state "v" in the sum-over-states in Eq. A20. Note that the contribution to the 3PA coefficient from a given initial valence band depends on the behavior of every other band due to the interference of 3PA pathways which pass through each band. However, for brevity we refer to these as the HH, LH, and SO contributions.

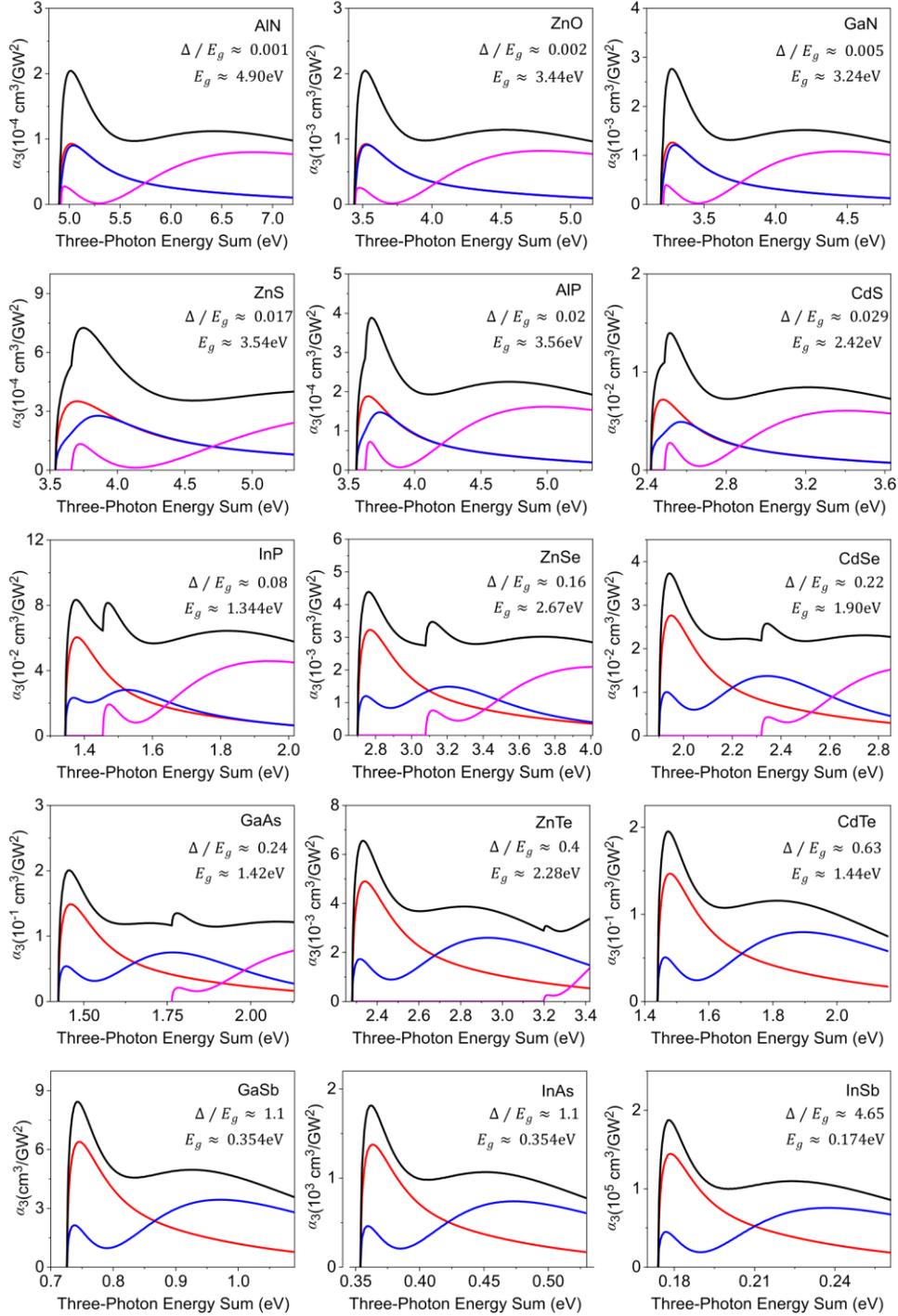

Fig. 5. The theoretical contributions of starting in the HH (red), LH (blue) and SO bands (pink) to the total (black) 3PA in several zinc-blende semiconductors using the 8-band model (see the appendices). These are ordered from small to large $\Delta/E_g$ to show the continuous evolution of the 3PA spectral shape, and the 3PA energy sum starts just below the onset of 3PA and goes to where 2PA turns on. No scaling is applied. Band parameters are summarized in Table 2.

The graphs of Fig 5, in order of increasing $\Delta/E_g$, show some general trends. We notice that the HH (red curves) and LH (blue curves) contributions to 3PA are nearly identical when $\Delta/E_g$ is very small, exemplified by AlN, ZnO, and GaN. 3PA coefficients from both bands exhibit a fast increase near a sum photon energy equal to the bandgap followed by a slow decrease back toward zero as the photon energy increases. The sharp shoulder can be attributed to the dominance of allowed-allowed-allowed transitions and interference effects are not readily apparent. In contrast, the SO contribution has the same shoulder but it is followed by a clear dip from destructive interference. As the photon energy increases, the SO contribution for $\Delta/E_g \geq 0.3$ starts to dominate so that the overall 3PA dispersion stays relatively flat through the 2PA edge.

As $\Delta/E_g$ increases, the LH and HH contributions develop different spectral shapes. The HH contribution maintains the shape it had with small spin-orbit coupling, while the LH contribution develops clear markers of destructive interference. The shape of the SO contribution also adjusts slightly as the additional influence from other quantum pathways reduces the net destructive interference. As seen in CdTe, GaSb, InAs, and InSb, there is no longer an SO contribution occurring below the 2PA edge when $\Delta/E_g > 0.5$.

In Fig. 6, we fix all band structure parameters except $\Delta$ and show the dependence of the normalized $\alpha_3$ on $\Delta/E_g$ to investigate the effects of spin-orbit coupling. As $\Delta/E_g$ decreases, the SO contribution becomes increasingly important in determining the peak 3PA coefficient and overall spectral shape. However, it is important to note that even where the SO contribution is small, the effect of spin-orbit coupling is still important in determining the shapes of the HH and LH contributions. The effect of spin-orbit interaction on 3PA is described more thoroughly in Appendix B. There, we also give more mathematical detail on the band structure features that give the HH, LH and SO contributions their spectral shape.

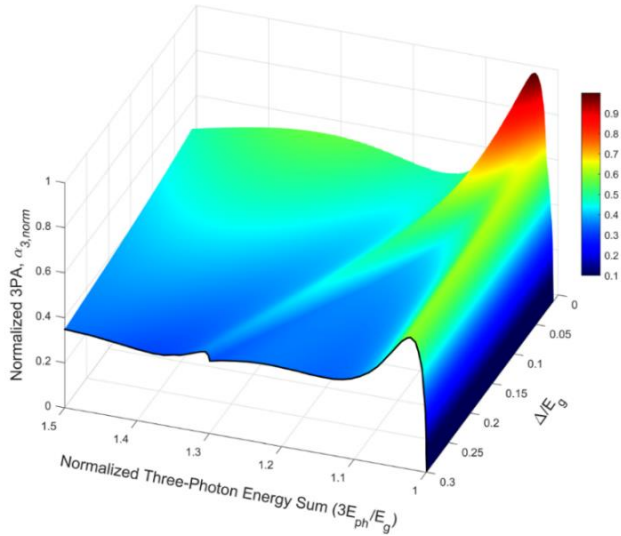

Fig. 6. Dependence of $\alpha_{3,norm}$ on the normalized three-photon energy sum and $\Delta/E_g$. $\alpha_{3,norm}$ is normalized to the value of 3PA when $\Delta = 0$. The range of $\Delta/E_g$ is limited to $\Delta/E_g < 0.3$ to better represent the effects of the SO band on the 3PA spectra.

## 6. CONCLUSIONS

Prior to this work, the theoretical 3PA bandgap and spectral scaling had not been satisfactorily confirmed by experiment, and conflicting results existed in the literature. We measured 3PA in nine semiconductors, confirming the expected bandgap scaling of $E_g^{-7}$ as predicted by simple 2-band theories as well as our 8-band theory. Eq. 2 (2-band theory of Wherrett [20]) gives the correct scaling and predictions within factors of a few for most of these materials but does not give the correct spectra. On the other hand, the 8-band analysis presented here accurately models 3PA spectra of ZnSe, ZnS, and GaAs as confirmed by our measurements. The agreement between our measured and calculated spectra shows the value of modeling with more sophisticated band structure; we are then able to predict with confidence the 3PA dispersions of 15 zinc-blende semiconductors for which little data exists. While this theory only strictly applies to zinc-blende crystals, we expect it to give reasonable predictions for materials with other symmetry such as $As_2Se_{1.5}S_{1.5}$ glass.

Our analysis, based on an 8-band model, separately yields the contributions to the 3PA spectra where the initial state is in the HH, LH, and SO bands. It is important to account for quantum interference between different pathways to predict the spectral dependence of 3PA, and the crucial importance of spin-orbit coupling in determining the spectral shape becomes apparent. The predictions made here can be extremely useful in the case of low-dimensional systems such as quantum wells, quantum dots, nanocrystals, and two-dimensional semiconductors where spin-orbit coupling and inter-subband transitions are present and may significantly alter the 3PA process via quantum interference between different bands [55-59].

## FUNDING

The authors gratefully acknowledge funding from the National Science Foundation, DMR-1609895.

## REFERENCES

1. M. Chattopadhyay, P. Kumbhakar, R. Sarkar, and A. Mitra, "Enhanced three-photon absorption and nonlinear refraction in ZnS and Mn 2+ doped ZnS quantum dots," Applied physics letters **95**, 163115 (2009).
2. G. S. He, J. D. Bhawalkar, P. N. Prasad, and B. A. Reinhardt, "Three-photon-absorption-induced fluorescence and optical limiting effects in an organic compound," Optics letters **20**, 1524-1526 (1995).
3. G. S. He, P. P. Markowicz, T.-C. Lin, and P. N. Prasad, "Observation of stimulated emission by direct three-photon excitation," Nature **415**, 767-770 (2002).
4. G. S. He, L.-S. Tan, Q. Zheng, and P. N. Prasad, "Multiphoton absorbing materials: molecular designs, characterizations, and applications," Chemical reviews **108**, 1245-1330 (2008).
5. J. He, Y. Qu, H. Li, J. Mi, and W. Ji, "Three-photon absorption in ZnO and ZnS crystals," Optics express **13**, 9235-9247 (2005).
6. W. C. Hurlbut, Y.-S. Lee, K. Vodopyanov, P. Kuo, and M. Fejer, "Multiphoton absorption and nonlinear refraction of GaAs in the mid-infrared," Optics letters **32**, 668-670 (2007).
7. S. Pearl, N. Rotenberg, and H. M. van Driel, "Three photon absorption in silicon for 2300–3300 nm," Applied Physics Letters **93**, 131102 (2008).
8. P. Zhao, M. Reichert, D. J. Hagan, and E. W. Van Stryland, "Dispersion of nondegenerate nonlinear refraction in semiconductors," Optics express **24**, 24907-24920 (2016).
9. G. Polónyi, B. Monoszlai, G. Gäumann, E. J. Rohwer, G. Andriukaitis, T. Balciunas, A. Pugzlys, A. Baltuska, T. Feurer, and J. Hebling, "High-energy terahertz pulses from semiconductors pumped beyond the three-photon absorption edge," Optics express **24**, 23872-23882 (2016).
10. M. R. Shcherbakov, K. Werner, Z. Fan, N. Talisa, E. Chowdhury, and G. Shvets, "Photon acceleration and tunable broadband harmonics generation in nonlinear time-dependent metasurfaces," Nature communications **10**, 1-9 (2019).
11. M. Reichert, A. L. Smirl, G. Salamo, D. J. Hagan, and E. W. Van Stryland, "Observation of nondegenerate two-photon gain in GaAs," Physical review letters **117**, 073602 (2016).
12. M. Reichert, P. Zhao, H. S. Pattanaik, D. J. Hagan, and E. W. Van Stryland, "Nondegenerate two-and three-photon nonlinearities in semiconductors," in *Ultrafast Bandgap Photonics*, (International Society for Optics and Photonics, 2016), 98350A.
13. J. He, W. Ji, J. Mi, Y. Zheng, and J. Y. Ying, "Three-photon absorption in water-soluble ZnS nanocrystals," Applied physics letters **88**, 181114 (2006).
14. A. D. Lad, P. Prem Kiran, G. Ravindra Kumar, and S. Mahamuni, "Three-photon absorption in ZnSe and Zn Se∕Zn S quantum dots," Applied physics letters **90**, 133113 (2007).
15. Y. Wang, V. D. Ta, Y. Gao, T. C. He, R. Chen, E. Mutlugun, H. V. Demir, and H. D. Sun, "Stimulated emission and lasing from CdSe/CdS/ZnS core-multi-shell quantum dots by simultaneous three-photon absorption," Advanced Materials **26**, 2954-2961 (2014).
16. S. S. Mitra, N. Judell, A. Vaidyanathan, and A. H. Guenther, "Three-photon absorption in direct-gap crystals," Optics letters **7**, 307-309 (1982).

17. Z. Wang, H. Liu, N. Huang, Q. Sun, J. Wen, and X. Li, "Influence of three-photon absorption on mid-infrared cross-phase modulation in silicon-on-sapphire waveguides," Optics express **21**, 1840-1848 (2013).
18. C. Husko, S. Combrié, Q. Tran, F. Raineri, A. De Rossi, and C. Wong, "Slow-light enhanced self-phase modulation, three-photon absorption and free-carriers in photonic crystals: experiment and theory," in *CLEO/QELS: 2010 Laser Science to Photonic Applications*, (IEEE, 2010), 1-2.
19. E. W. Van Stryland, H. Vanherzeele, M. A. Woodall, M. Soileau, A. L. Smirl, S. Guha, and T. F. Boggess, "Two photon absorption, nonlinear refraction, and optical limiting in semiconductors," Optical Engineering **24**, 244613 (1985).
20. B. Wherrett, "Scaling rules for multiphoton interband absorption in semiconductors," JOSA B **1**, 67-72 (1984).
21. E. W. Van Stryland, M. Woodall, H. Vanherzeele, and M. Soileau, "Energy band-gap dependence of two-photon absorption," Optics letters **10**, 490-492 (1985).
22. J. H. Yee, "Three-photon absorption in semiconductors," Physical Review B **5**, 449 (1972).
23. A. Vaidyanathan, T. Walker, A. Guenther, S. Mitra, and L. Narducci, "Two-photon absorption in several direct-gap crystals," Physical Review B **21**, 743 (1980).
24. A. Vaidyanathan, A. Guenther, and S. Mitra, "Two-photon absorption in direct-gap crystals—an addendum," in *Laser Induced Damage In Optical Materials: 1980* (ASTM International, 1981).
25. J. M. Hales, S.-H. Chi, T. Allen, S. Benis, N. Munera, J. W. Perry, D. McMorrow, D. J. Hagan, and E. W. Van Stryland, "Third-order nonlinear optical coefficients of Si and GaAs in the near-infrared spectral region," in *CLEO: Applications and Technology*, (Optical Society of America, 2018), JTu2A. 59.
26. L. V. Keldysh, "Zh. Éksp. Teor. Fiz. 47, 1945 1964 Sov. Phys.," Jetp **20**, 1307 (1965).
27. C. M. Cirloganu, P. D. Olszak, L. A. Padilha, S. Webster, D. J. Hagan, and E. W. Van Stryland, "Three-photon absorption spectra of zinc blende semiconductors: theory and experiment," Optics letters **33**, 2626-2628 (2008).
28. C. M. Cirloganu, P. D. Olszak, L. A. Padilha, S. Webster, D. J. Hagan, and E. W. Van Stryland, "Three-photon absorption spectra of zinc blende semiconductors: theory and experiment: erratum," Optics Letters **45**, 1025-1026 (2020).
29. H. Brandi and C. De Araujos, "Multiphonon absorption coefficients in solids: a universal curve," Journal of Physics C: Solid State Physics **16**, 5929 (1983).
30. E. O. Kane, "Band structure of indium antimonide," Journal of Physics and Chemistry of Solids **1**, 249-261 (1957).
31. M. Hasselbeck, A. Said, E. Van Stryland, and M. Sheik-Bahae, "Three-photon absorption in InAs," Optical and quantum electronics **30**, 193-200 (1998).
32. J. U. Kang, A. Villeneuve, M. Sheik-Bahae, G. I. Stegeman, K. Al-hemyari, J. S. Aitchison, and C. N. Ironside, "Limitation due to three-photon absorption on the useful spectral range for nonlinear optics in AlGaAs below half band gap," Applied physics letters **65**, 147-149 (1994).
33. G.-M. Schucan, R. G. Ispasoiu, A. M. Fox, and J. F. Ryan, "Ultrafast two-photon nonlinearities in CdSe near 1.5/spl mu/m studied by interferometric autocorrelation," IEEE journal of quantum electronics **34**, 1374-1379 (1998).
34. M. Sheik-Bahaei, P. Mukherjee, and H.-S. Kwok, "Two-photon and three-photon absorption coefficients of InSb," JOSA B **3**, 379-385 (1986).
35. J. Bechtel and W. L. Smith, "Two-photon absorption in semiconductors with picosecond laser pulses," Physical Review B **13**, 3515 (1976).
36. M. Sheik-Bahae, A. A. Said, T.-H. Wei, D. J. Hagan, and E. W. Van Stryland, "Sensitive measurement of optical nonlinearities using a single beam," IEEE journal of quantum electronics **26**, 760-769 (1990).
37. M. Sheik-Bahae, A. A. Said, and E. W. Van Stryland, "High-sensitivity, single-beam n 2 measurements," Optics letters **14**, 955-957 (1989).
38. S. Shabahang, G. Tao, M. P. Marquez, H. Hu, T. R. Ensley, P. J. Delfyett, and A. F. Abouraddy, "Nonlinear characterization of robust multimaterial chalcogenide nanotapers for infrared supercontinuum generation," JOSA B **31**, 450-457 (2014).
39. M. A. Woodall, "Nonlinear absorption techniques and measurements in semiconductors," (1985).
40. S. Guha, E. W. Van Stryland, and M. Soileau, "Self-defocusing in CdSe induced by charge carriers created by two-photon absorption: erratum," Optics letters **11**, 59-59 (1986).
41. S. Guha, E. W. Van Stryland, and M. Soileau, "Self-defocusing in CdSe induced by charge carriers created by two-photon absorption," Optics letters **10**, 285-287 (1985).
42. P. D. Olszak, C. M. Cirloganu, S. Webster, L. A. Padilha, S. Guha, L. P. Gonzalez, S. Krishnamurthy, D. J. Hagan, and E. W. Van Stryland, "Spectral and temperature dependence of two-photon and free-carrier absorption in InSb," Physical Review B **82**, 235207 (2010).
43. S. Z. Karazhanov and L. L. Y. Voon, "Ab initio studies of the band parameters of III–V and II–VI zinc-blende semiconductors," Semiconductors **39**, 161-173 (2005).
44. S. Krishnamurthy, Z. G. Yu, L. P. Gonzalez, and S. Guha, "Temperature-and wavelength-dependent two-photon and free-carrier absorption in GaAs, InP, GaInAs, and InAsP," Journal of Applied Physics **109**, 033102 (2011).

45. I. Vurgaftman, J. á. Meyer, and L. á. Ram-Mohan, "Band parameters for III–V compound semiconductors and their alloys," Journal of applied physics **89**, 5815-5875 (2001).
46. B. V. Olson, M. P. Gehlsen, and T. F. Boggess, "Nondegenerate two-photon absorption in GaSb," Optics Communications **304**, 54-57 (2013).
47. J. Wei, J. Murray, C. Reyner, and S. Guha, "Measurement of Wavelength and Temperature Dependent Refractive Index of GaSb," in *Novel Optical Materials and Applications*, (Optical Society of America, 2019), NoM3B. 3.
48. M. Cardona, K. L. Shaklee, and F. H. Pollak, "Electroreflectance at a semiconductor-electrolyte interface," Physical Review **154**, 696 (1967).
49. P. Rochon and E. Fortin, "Photovoltaic effect and interband magneto-optical transitions in InP," Physical Review B **12**, 5803 (1975).
50. P. Zory Jr, "Quantum Well Lasers (Academic Press, 1993)."
51. M. Cardona and G. Harbeke, "Optical properties and band structure of wurtzite-type crystals and rutile," Physical Review **137**, A1467 (1965).
52. M. Willatzen, M. Cardona, and N. Christensen, "Spin-orbit coupling parameters and electron g factor of II-VI zinc-blende materials," Physical Review B **51**, 17992 (1995).
53. D. Hutchings and E. W. Van Stryland, "Nondegenerate two-photon absorption in zinc blende semiconductors," JOSA B **9**, 2065-2074 (1992).
54. D. Thomas, "The exciton spectrum of zinc oxide," Journal of Physics and Chemistry of Solids **15**, 86-96 (1960).
55. Y. S. Ang and C. Zhang, "Step-like multi-photon absorption in two-dimensional semiconductors with Rashba spin-orbit coupling in terahertz regime," in *2014 39th International Conference on Infrared, Millimeter, and Terahertz waves (IRMMW-THz)*, (IEEE, 2014), 1-1.
56. X. Feng, Y. L. Ang, J. He, C. W. Beh, H. Xu, W. S. Chin, and W. Ji, "Three-photon absorption in semiconductor quantum dots: experiment," Optics express **16**, 6999-7005 (2008).
57. A. Mang and K. Reimann, "Band gaps, crystal-field splitting, spin-orbit coupling, and exciton binding energies in ZnO under hydrostatic pressure," Solid state communications **94**, 251-254 (1995).
58. P. Lambropoulos and M. Teague, "Two-photon ionization with spin-orbit coupling," Journal of Physics B: Atomic and Molecular Physics **9**, 587 (1976).
59. J. H. Davies, *The physics of low-dimensional semiconductors: an introduction* (Cambridge university press, 1998).
60. S. L. Chuang and S. L. Chuang, "Physics of optoelectronic devices," (1995).
61. S. M. Sze and K. K. Ng, *Physics of semiconductor devices* (John wiley & sons, 2006).
62. M. Dinu, "Dispersion of phonon-assisted nonresonant third-order nonlinearities," IEEE journal of quantum electronics **39**, 1498-1503 (2003).
63. R. W. Boyd, *Nonlinear optics* (Elsevier, 2003).

## APPENDIX A: THEORETICAL APPROACH FOR CALCULATING 3PA USING THE KANE 8-BAND MODEL

In this appendix, we provide detail about the theoretical treatment of 3PA in zinc-blende semiconductors. The first subsection focuses on the Kane 8-band model for electronic states and the second part combines the band structure with third order perturbation theory to calculate 3PA coefficients. Note that we refer to the model as comprising 8 bands, while it is sometimes described in the literature as a 4-band model. The reason for this becomes apparent in the second part of the appendix; although a given pair of bands may be energy degenerate, they each have their own optical selection rules and therefore must be treated as separate entities.

*Kane band structure*

We use the band structure model developed by Kane [30] because it includes band symmetry intermixing that is ignored in simplified 2-band models. We re-derive the model here for reference following the original work and the excellent summary in Ref. [60]. We start with the Schrödinger equation in the following form:

$$H_0 \Psi_n(\mathbf{r}) = E_n \Psi(\mathbf{r}), \quad (A1)$$

where $H_0 = \mathbf{p}^2/(2m_0) + V(r)$ is the usual Hamiltonian operator. $\Psi_n$ and $E_n$ are the wavefunction and energy, respectively, of an electron in an eigenstate labeled by $n$. The symbol

$\mathbf{p}$ is the momentum operator, $m_0$ is the electron mass, and $V(\mathbf{r})$ is the one-electron potential. Substituting the Bloch form,

$$\Psi_{n\mathbf{k}}(\mathbf{r}) = \exp(i\mathbf{k} \cdot \mathbf{r}) u_{n\mathbf{k}}(\mathbf{r}), \tag{A2}$$

into Eq. A1 gives the Hamiltonian

$$H_0 u_{n\mathbf{k}}(\mathbf{r}) + \left(\frac{\hbar}{m}\right) \mathbf{k} \cdot \mathbf{p}\, u_{n\mathbf{k}}(\mathbf{r}) = E'_n u_{n\mathbf{k}}(\mathbf{r}), \tag{A3}$$

where we have defined

$$E'_n = E_n - \frac{\hbar^2 k^2}{2m_0}. \tag{A4}$$

At the zone center ($\mathbf{k} = 0$), Eq. A3 reduces to a bare Schrodinger equation for each unit cell function $u_{n0}$. We choose a finite set of orthonormal $u_{n0}$ and assume that they form a complete set, then expand the $u_{n\mathbf{k}}$ of Eq. A3 in this basis as

$$u_{n\mathbf{k}} = \sum_m c_{mn}(\mathbf{k}) u_{m0}(\mathbf{r}). \tag{A5}$$

Note that we could have chosen a basis defined anywhere throughout the Brillouin zone, but the zone center is the most convenient reference because it is the point of highest symmetry. Obviously, the accuracy of the expansion in Eq. A5 depends on the dimensionality of the basis. The bands mix as the wave vector increases due to interaction from the $\mathbf{k} \cdot \mathbf{p}$ term, which is the origin of the term $\mathbf{k} \cdot \mathbf{p}$ theory. By pre-multiplying $u_{k0}(\mathbf{r})$ and integrating over the crystal volume, we can form a matrix equation for the expansion coefficients $c_{kn}$. This matrix equation can be solved perturbatively or diagonalized exactly. Here, we diagonalize the matrix exactly and ignore perturbative contributions from bands outside the basis.

We include spin by taking the total unit cell function to be the tensor product of spatial and spin degrees of freedom, then the spin-orbit interaction term

$$H_{\text{SO}} = \frac{\hbar}{4m_0^2 c^2} (\nabla V \times \mathbf{p}) \cdot \boldsymbol{\sigma} \tag{A6}$$

is added to Eq. A1. In Eq. A6, $\boldsymbol{\sigma}$ is the vector of Pauli spin matrices that act on the spin components and $c$ is the speed of light. Substituting the Bloch form (Eq. A2) into the updated Hamiltonian simplifies to

$$\left[H_0 + \left(\frac{\hbar}{m}\right) \mathbf{k} \cdot \mathbf{p} + \left(\frac{\hbar}{4m^2 c^2}\right)(\nabla V \times \mathbf{p}) \cdot \boldsymbol{\sigma}\right] |u_{n\mathbf{k}}(\mathbf{r}); s\rangle = E'_n |u_{n\mathbf{k}}(\mathbf{r}); s\rangle \tag{A7}$$

after neglecting $\mathbf{k}$-dependent spin-orbit terms. In Eq. A7 we separated the spatial and spin ($s$) degrees of freedom with a semicolon, but henceforth we will represent the spin by an arrow after the spatial component's label.

In the zinc-blende structures studied here, the states closest to the Fermi energy are a doubly spin degenerate $s$-like conduction band and $p$-like valence bands that are six-fold degenerate (including spin) before spin-orbit interaction. While the 8 basis functions $|iS \uparrow\rangle$, $|iS \downarrow\rangle$, $|X \uparrow\rangle$, $|X \downarrow\rangle$, $|Y \uparrow\rangle$, $|Y \downarrow\rangle$, $|Z \uparrow\rangle$, and $|Z \downarrow\rangle$ are a perfectly acceptable choice, we find the most convenient Hamiltonian matrix by selecting valence basis functions from the spherical harmonics $Y_{10} = |Z\rangle$ and $Y_{1\pm1} = \mp(X \pm iY)/\sqrt{2}$. Thus, the full ordered basis is taken to be

$$|iS \downarrow\rangle, \left|\frac{(X - iY)}{\sqrt{2}} \uparrow\right\rangle, |Z \downarrow\rangle, \left|\frac{(X + iY)}{\sqrt{2}} \uparrow\right\rangle, \tag{A8}$$

$$|iS\uparrow\rangle, \left|-\frac{(X+iY)}{\sqrt{2}}\downarrow\right\rangle, |Z\uparrow\rangle, \left|\frac{(X-iY)}{\sqrt{2}}\downarrow\right\rangle.$$

Fixing $\mathbf{k} = k\hat{\mathbf{z}}$, the Hamiltonian matrix is in block diagonal form

$$\begin{bmatrix} H & 0 \\ 0 & H \end{bmatrix} \text{ where } H = \begin{bmatrix} E_s & 0 & kP & 0 \\ 0 & E_p - \frac{\Delta}{3} & \frac{\sqrt{2}\Delta}{3} & 0 \\ kP & \frac{\sqrt{2}\Delta}{3} & E_p & 0 \\ 0 & 0 & 0 & E_p + \frac{\Delta}{3} \end{bmatrix}. \tag{A9}$$

$E_s$ and $E_p$ are the zone center energies defined by $H_0 u_{j0} = E_j u_{j0}$. These are chosen to be $E_s = E_g$ and $E_p = -\Delta/3$ so that the last row and column go to zero. This choice reduces the problem to the diagonalization of two identical $3 \times 3$ matrices. $P$ is the Kane parameter defined by

$$P = -i(\hbar/m)\langle S|p_z|Z\rangle = -i(\hbar/m)\langle S|p_y|Y\rangle = -i(\hbar/m)\langle S|p_x|X\rangle \tag{A10}$$

and $\Delta$ is the spin-orbit interaction energy

$$\Delta = \frac{3\hbar i}{4m^2 c^2}\left\langle X\left|\frac{\partial V}{\partial x}p_y - \frac{\partial V}{\partial y}p_x\right|Y\right\rangle. \tag{A11}$$

All equivalent matrix elements can be generated from Eq. A11 by applying all symmetry operations of the crystal's point group, as was done for $P$ in Eq. A10.

Diagonalizing Eq. A9 at the zone center gives eigenstates of the total angular momentum $J = L + S$ (orbital plus spin) and its z component $J_z$. The s-like ($L = 0$) conduction band is a doubly degenerate state with $L = 0$ and $S = \pm 1/2$. The p-like ($L = 1$) valence bands include heavy holes with quantum numbers $(J, J_z) = (3/2, \pm 3/2)$, light holes with $(3/2, \pm 1/2)$, and split-off holes with $(1/2, \pm 1/2)$. Spin-orbit coupling serves to split the $J = 1/2$ states from $J = 3/2$ states by energy $\Delta$, hence the name split-off holes.

For arbitrary $\mathbf{k}$, we simply rotate to primed coordinates where $\mathbf{k} = k\hat{\mathbf{z}}'$. The Hamiltonian in the new coordinates has the exact same form as Eq. A7, but with basis functions transformed according to their representation. Because the valence bands belong to $\Gamma_4$ which transform as the components of a vector, the transformation takes the form of the usual vector rotation matrix

$$\begin{bmatrix} X' \\ Y' \\ Z' \end{bmatrix} = \begin{bmatrix} \cos\theta\cos\phi & \cos\theta\sin\phi & -\sin\theta \\ -\sin\phi & \cos\phi & 0 \\ \sin\theta\cos\phi & \sin\theta\sin\phi & \cos\theta \end{bmatrix} \begin{bmatrix} X \\ Y \\ Z \end{bmatrix}. \tag{A12}$$

The spin components transform as the 2-dimensional irreducible representation of SU(2),

$$\begin{bmatrix} \uparrow' \\ \downarrow' \end{bmatrix} = \begin{bmatrix} e^{-\phi/2}\cos(\theta/2) & e^{\phi/2}\sin(\theta/2) \\ e^{-\phi/2}\sin(\theta/2) & e^{\phi/2}\cos(\theta/2) \end{bmatrix} \begin{bmatrix} \uparrow \\ \downarrow \end{bmatrix}, \tag{A13}$$

but this rotation is not necessary to compute since the optical selection rule remains unchanged by rotation. As the conduction bands belong to the identity (spherical) representation $\Gamma_1$, $S' = S$ is invariant under rotation. The angles $\theta$ and $\phi$ are the usual polar angles of the $\mathbf{k}$ vector relative to the crystal axes labeled $x, y$ and $z$. The characteristic equation yields four double roots corresponding to the eigenvalues of Eq. A9:

$$\begin{aligned} E' &= 0 \\ E'(E' - E_G)(E' + \Delta) &- k^2 P^2(E' + 2\Delta/3) = 0, \end{aligned} \tag{A14}$$

where $E_k = E'_k + (\hbar^2/2m_0)k^2$ is the energy of a state with a wavevector **k**. The electronic wave functions are found to be

$$
\begin{aligned}
u_{i\alpha} &= a_i[iS\downarrow]' + b_i\big[(X-iY)\uparrow/\sqrt{2}\big]' + c_i[Z\downarrow]' \\
u_{i\beta} &= a_i[iS\uparrow]' + b_i\big[-(X+iY)\downarrow/\sqrt{2}\big]' + c_i[Z\uparrow]' \\
u_{HH\alpha} &= \big[(X+iY)\uparrow/\sqrt{2}\big]' \\
u_{HH\beta} &= \big[(X-iY)\downarrow/\sqrt{2}\big]',
\end{aligned}
\tag{A15}
$$

where $\alpha$ and $\beta$ denote the two degenerate spin states and the index $i$ refers to the conduction, LH, and SO bands. The real coefficients $a, b$, and $c$ are obtained from

$$
\begin{aligned}
a_i &= kP(E'_i + 2\Delta/3)/N \\
b_i &= (\sqrt{2}\Delta/3)(E'_i - E_G)/N \\
c_i &= (E'_i - E_G)(E'_i + 2\Delta/3)/N
\end{aligned}
\tag{A16}
$$

where $N$ is a normalizing factor such that $a_i^2 + b_i^2 + c_i^2 = 1$.

## 3PA COEFFICIENTS

Here, we use the results of the previous subsection to compute 3PA coefficients in third order perturbation theory. We begin with the electron-radiation interaction Hamiltonian in SI units [61, 62]

$$
H_p = \frac{e}{i\omega m_0}\left(\frac{I}{2\varepsilon_0 n_0 c}\right)^{1/2} \hat{\mathbf{a}}\cdot\hat{\mathbf{p}},
\tag{A17}
$$

where $I$ is the irradiance of the incident beam, $n_0$ is the material's index of refraction and $\hat{\mathbf{a}}$ is a unit vector parallel to the direction of the incident electric field. We can then express the total 3PA transition rate per unit volume using Fermi's Golden rule derived from third-order perturbation theory:

$$
W_3 = \frac{2\pi}{V\hbar}\sum_{c,v}\sum_{\mathbf{k}}\left|\sum_{i,j}\frac{\langle c|H_p|j\rangle'\langle j|H_p|i\rangle'\langle i|H_p|v\rangle'}{(E_{jv}(\mathbf{k}) - 2\hbar\omega)(E_{iv}(\mathbf{k}) - \hbar\omega)}\right|^2 \delta[E_{cv}(\mathbf{k}) - 3\hbar\omega],
\tag{A18}
$$

with $E_{mn} := E_m - E_n$. The index $v$ corresponds to a valence band (initial state), $c$ is a conduction band (final state), and $i$ and $j$ are intermediate states chosen from any of the eight bands detailed in the previous subsection. The kets $|l\rangle$ (and bras $\langle l|$) are shorthand for the **k**-dependent unit cell functions $u_{l\mathbf{k}}$ as calculated by Eqs. A15 and A16; the plane wave portion of the Bloch state (Eq. A2) has already been used to enforce the crystal momentum selection rule leaving a single sum over wave vector **k**. Primes on the interaction matrix elements indicate that basis functions are rotated according to the transformation in Eq. A12. Note that quantum interference occurs due to the intermediate state sum within the absolute value brackets for a given $c, v$ and **k**. Since each term in this sum can have arbitrary sign and magnitude, the net contribution to 3PA can only be determined by adding up the effect of all possible paths.

The transition rate per unit volume is used to calculate the 3PA coefficient by relating electron transition rate to photon flux with [20, 63]:

$$
\alpha_3(\omega) = \frac{3\hbar\omega W_3}{I^3}.
\tag{A19}
$$

Combining Eqs. A17, A18, and A19 yields the final expression for 3PA coefficients for z-polarized light:

$$\alpha_3 = 3\omega \frac{(2\pi)^5}{(n_0 c)^3} \left(\frac{eP}{\hbar\omega}\right)^6 \sum_{c,v} \int_0^\pi \left( \left| \sum_{i,j} \frac{M_{cj}^z(k_r,\theta) M_{ji}^z(k_r,\theta) M_{iv}^z(k_r,\theta)}{(E_{iv}(k_r) - \hbar\omega)(E_{jv}(k_r) - 2\hbar\omega)} \right|^2 \right) \frac{k_r^2 \sin\theta \, d\theta}{\left|\frac{\partial E_{cv}(k)}{\partial k}\right|_{k=k_r}}, \quad (A20)$$

where

$$M_{ij}^z(k_r,\theta) = 3\frac{\hbar}{m_0 P} \langle u_i(k_r,\theta) | p_z | u_j(k_r,\theta) \rangle' \quad (A21)$$

are the normalized momentum matrix elements [53]. Note that the **k** sum in Eq. A18 has been converted to a volume integral by $(1/V)\sum_k \to \int (2\pi)^3 d^3\mathbf{k}$ and evaluated in spherical coordinates. The remaining $\delta$-function integration results in an integral over constant energy surfaces defined by

$$E_{cv}(k_r) - 3\hbar\omega = 0, \quad (A22)$$

with $k_r$ denoting the $k$ value for which the transition is resonant.

The z-direction was chosen so that matrix elements are independent of azimuthal angle. Inspection of the third column of the rotation matrix Eq. A12 shows that the $Z$ component of rotated basis functions only depends on angle $\theta$. Since the only nonzero matrix element for z-polarized light is of the form $\langle iS | p_z | Z \rangle$, the matrix elements only depend on $\theta$ while the $\phi$ integral evaluates to $2\pi$. We chose z polarization for convenience, but analysis for light polarized in any direction would yield the same result due to the spherical symmetry of the Kane band structure.

## APPENDIX B: SPIN-ORBIT COUPLING AND QUANTUM INTEREFRENCE IN 3PA SPECTRA

This appendix aims to provide insight into the shapes and parameter dependence of the curves in Fig. 5. Given the large number of possible 3PA pathways, it is not immediately obvious which terms dominate at a given three-photon energy sum. While it is possible to write each term individually and compare the magnitudes at a given point in $k$-space, it is just as illuminating to give a semi-qualitative reason for the dispersions shown. We begin by carefully examining the case with no spin-orbit coupling, and demonstrate how the complexity in the spectrum increases with this coupling.

The conceptually simplest description exists for materials with weak spin-orbit coupling: AlN, ZnO, and GaN. We study these by assuming $\Delta = 0$ so that only $\mathbf{k}\cdot\mathbf{p}$ coupling terms between $|S\rangle'$ and $|Z\rangle'$ remain in the Hamiltonian matrix (Eq. A9). We are free to rearrange the order of the basis to group functions of the same spin into a single block. Without loss of generality, we diagonalize the spin up block to find

$$\begin{aligned} u_C &= a_c(k)|iS\rangle' + b_B(k)|Z\rangle' \\ u_{SO} &= a_z(k)|iS\rangle' + b_z(k)|Z\rangle' \\ u_{HH} &= \frac{1}{\sqrt{2}}|X - iY\rangle' \\ u_{LH} &= \frac{1}{\sqrt{2}}|X + iY\rangle'. \end{aligned} \quad (A23)$$

The conduction and SO bands intermix while the HH and LH bands remain uncoupled and retain a free-electron-like dispersion. We compute $z$ matrix elements by rotating the primed

basis back to unprimed functions using Eq. A12. Then applying Eq. A10 and normalizing by Eq. A21 yields

$$M^z_{C-\text{HH}} = \langle u_c|p_z|u_{\text{HH}}\rangle = \frac{3}{\sqrt{2}} a_c \sin\theta$$

$$M^z_{C-\text{LH}} = \langle u_c|p_z|u_{\text{LH}}\rangle = \frac{3}{\sqrt{2}} a_c \sin\theta.$$

(A24)

The matrix elements are identical because $\langle iS|p_z|Z\rangle$ is the only nonzero matrix element by symmetry, and only the $|X\rangle'$ term contains a $|Z\rangle$ component when rotated into the unprimed coordinates. The same behavior exists for $M_{c-\text{so}}$; the $|Z\rangle$ couples with the $|iS\rangle$ component of $u_{\text{so}}$. Because these bands experience the same optical coupling and are energy degenerate throughout the Brillouin zone, contributions to the 3PA coefficient from transitions originating in LH and HH bands will be identical. As we mentioned in the main text, for brevity we refer to these as the HH and LH contributions. By comparing the red and blue curves in Fig. 5 for AlN, ZnO, and GaN, we can see the equality of these contributions except for minor difference because $\Delta$ is not exactly zero.

We can reasonably approximate the shape of these curves with low spin-orbit coupling by the expression

$$\alpha_3 \propto \frac{1}{x^5}\left|\frac{A}{x^2} + \frac{B}{x^2}(3x-1)\right|^2 (3x-1)^{\frac{1}{2}},$$

(A25)

where $x$ is the photon energy normalized to the bandgap, $\hbar\omega/E_g$. Eq. A25 is derived from Eq. A20 by first solving for $k_r$ using the approximate energy conservation condition

$$E_{cv}(k_r) - 3\hbar\omega \approx E_g + \frac{\hbar^2 k_r^2}{2\mu_{cv}} - 3\hbar\omega = 0,$$

(A26)

where $1/\mu_{cv} = (1/\mu_c - 1/\mu_v)$ is the inverse reduced effective mass. Normalizing all energies, we obtain

$$k_r = \frac{\sqrt{2\mu_{cv}}}{\hbar}(3x-1)^{1/2}.$$

(A27)

The $A$ term in Eq. A25 describes triply-allowed 3PA pathways with **k**-independent matrix elements. The $B$ term accounts for allowed-forbidden-forbidden transitions and picks up a $k_r^2$ term because each forbidden matrix element is approximately linear in $k$ near the zone center.

Quantum interference arises due to differing relative magnitudes and signs between A and B in Eq. A25. The triply-allowed term dominates for HH and LH contributions when $\Delta \approx 0$, leading to a 3PA which scales as Wherrett's theory predicts (Eq. 2). From this, we can determine that there is either constructive or weak destructive interference and the actual result can only be found by tediously expanding each term in the sum and comparing. In contrast to the HH and LH, the SO contribution clearly demonstrates strong deconstructive interference: the triply-allowed term still dominates near the zone center, but the allowed-forbidden-forbidden terms have opposite sign and grow as $k_r$ increases. Eventually this term grows larger than the triply-allowed term, flipping the sign of the sum. Since the 3PA coefficient scales with the magnitude squared of the sum, this sign change appears as a drop down to zero followed by a bounce back upward. This is very clearly seen in the pink curve for ZnO.

As spin-orbit splitting increases from zero, the LH wave function becomes distinguished from that of the HH. However, this identity only persists throughout a region of the Brillouin zone determined by the strength of the spin-orbit interaction. Quantitatively, this means that $a_{\text{LH}}$ and $c_{\text{LH}}$ of Eq. A15 decay away to zero as $k$ increases so that the wave function tends to the same form as that for HH. When spin-orbit coupling is small, this decrease occurs very quickly. See Fig. A1 for a plot of the expansion coefficients: A1(a) shows how the wave functions change throughout the Brillouin zone in InP ($\Delta/E_g = 0.08$), and A1(b) shows what happens when $\Delta/E_g = 0.008$, near to the value for GaN. Figure A2(a) shows the band structure obtained for InP. To facilitate easy comparison with features in Fig. 5, we also plot in Fig. A2(b) the same band structure but with the horizontal axis to be the three-photon energy sum resonant with light hole to conduction transitions. Specifically, three-photon energy sums correspond to a given $k$ by the relation $E_{cv}(k) = 3\hbar\omega$ where $v$ = LH. Fig. A2(a) shows that the slopes of HH and LH band energies become nearly identical at large $k$.

For intermediate $\Delta/E_g$, we see that the LH and HH contributions in Fig. 5 nearly equalize at some photon energy. Some minor differences may remain, attributable to differing denominators in the perturbation expansion. This merging of LH and HH contributions is apparent in the 3PA curves for InP, AlP, ZnS and CdS. Comparing Fig. A1 and Fig. 5, we see that HH and LH 3PA contributions approach each other as the magnitude of the HH-like expansion coefficient ($|b_{LH}|$) of the LH band nears unity.

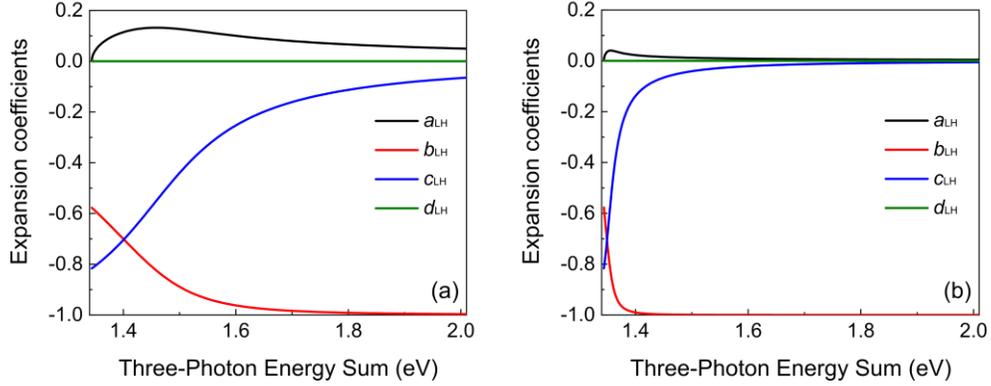

Fig. A1. Expansion coefficients for the light hole unit cell function versus three-photon energy sum for simulated values of (a) $\Delta/E_g = 0.08$ and (b) $\Delta/E_g = 0.008$. The horizontal axis is related to $k$ by $E_{cv}(k) = 3\hbar\omega$ for $v = LH$.

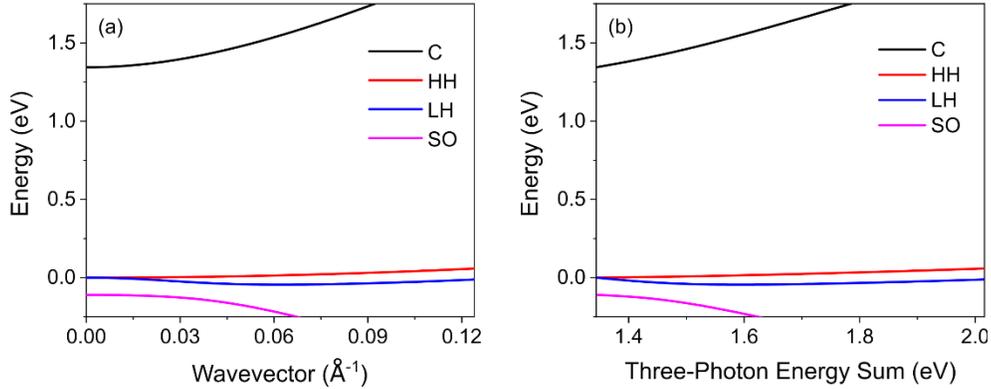

Fig. A2. Energy bands of InP calculated from parameters given in Table 2 with horizontal axes chosen to be (a) $k$ and (b) three-photon energy sum resonant with the conduction (C) to LH gap at a given $k$ ($E_{cv}(k) = 3\hbar\omega$ with $v = LH$). The ranges of $k$ values in (a) and (b) are identical.

LH contributions for intermediate $\Delta/E_g$ still experience destructive interference as the triply-allowed term is partially cancelled by allowed-forbidden-forbidden transitions. Note, however, that this cancellation is more complex than the that of SO holes with $\Delta = 0$. As shown in Fig.

A1, the $|iS\rangle$ component of the LH wavefunction increases then decreases, leading to non-monotonic forbidden transition matrix elements that depend on $a_{\text{LH}}$. This non-monotonicity partially explains why the interference does not lead the 3PA coefficient to pass through 0 as in the $\Delta = 0$ limit. For materials with large spin-orbit coupling, the LH and HH curves remain distinct throughout the entire region of interest and the 3PA spectral shape is not easily attributed to a specific interaction term. This is the behavior exhibited by CdTe, GaSb, InAs, and InSb.